\documentclass{article}

\usepackage[utf8]{inputenc}
\usepackage{amsmath}
\usepackage{amssymb}
\usepackage{theorem}

\newtheorem{Def}{Definition}
\newtheorem{Theo}{Theorem}
\newtheorem{Prop}[Theo]{Proposition}
\newtheorem{Lemma}[Theo]{Lemma}
\newtheorem{Cor}{Corollary}[Theo]

\newenvironment{Proof}[1][Proof]{\paragraph{{#1}}}%
                {{\hfill\(\Box\)\\}}
                {{\hfill\(\Box\)\\}}
        {\paragraph{{#1}}\begin{list}{}{}\item
        }{\end{list}}

\newcommand{\cand}{\text{ and }}
\newcommand{\cor}{\text{ or }}

\newcommand{\coll}[1]{\ensuremath{\left\{ {#1}\right\} }}
\newcommand{\pair}[2]{\ensuremath{\left\langle {#1}, {#2} \right\rangle}}

\newcommand{\paren}[1]{\ensuremath{\left( {#1} \right)}}

\newcommand{\set}[2]{\ensuremath{\left\{\left.#1\,\,\vphantom{#2}\right|\,#2\right\}}}
\newcommand{\fall}[1]{{\forall\,{#1},\ }}
\newcommand{\fexist}[1]{{\exists\,{#1}\,{:}\ }}

\newcommand{\mc}[1]{{\mathcal{#1}}}

\newcommand{\mb}[1]{{\bf #1}}

\title{An Intrisic Topology for Orthomodular Lattices}
\author{Olivier Brunet\footnote{olibrunet@free.fr}}

\newcommand{\sas}{\,\&\,}

\begin{document}

\maketitle

\begin{abstract}
We present a general way to define a topology on orthomodular lattices. We show that in the case of a Hilbert lattice, this topology is equivalent to that induced by the metrics of the corresponding Hilbert space. Moreover, we show that in the case of a boolean algebra, the obtained topology is the discrete one. Thus, our construction provides a general tool for studying orthomodular lattices but also a way to distinguish classical and quantum logics.
\end{abstract}

\section{Introduction}

The study of quantum structures (also called {\em quantum logics} \cite{DallaChiara2001QuantumLogic}) relates to algebraic order-theoretic structures (such as orthomodular lattices, orthomodular posets, orthoalgebras \cite{Foulis92Filters}, etc.) abstracted from Hilbert lattices, that is the collection of closed subspaces (or equivalently, of projection operators) of a Hilbert space. But Hilbert lattices are also much more than just order-theoretic objects, since they possesses a rich topological structure. It is thus reasonable and tempting to study the possibility of defining topologies on more general quantum structures.

There already exists some litterature concerning topology and quantum structures. One can mention study of orthomodular lattices which are also topological lattices \cite{choe94Orthotopo,Greechie93PMOL,Pulman93BFOML} or of topological orthoalgebras \cite{Wilce05AMS,wilce2005b,wilce2005IJTP}. However, there does not exists general results concerning the possibility of defining a topology on orthomodular lattices. This remark motivates the present article.

In the next section, we present a geometric lemma which relates the metrics of the projective sphere of a Hilbert lattice to lattice-theoretical considerations. We then show how this result can be used to define a topology, first on some particular atomic orthomodular lattices and then on any orthomodular lattice. We also show that the obtained general construction permits to recover the metric-induced topology in the case of a Hilbert lattice. Finally, we prove that in the case of a boolean algebra, this construction leads to the discrete topology. As a result, our topological construction provides, together with a general tool for studying orthomodular lattices, a way to discriminate classical boolean structures from some interesting quantum structures.

\section{A Mathematical Preliminary}\label{sec:a_mathematical_preliminary}

We first show how it is possible to recover the metrics of the projective sphere of a Hilbert lattice in a purely lattice-theoretical way. This result is based on a geometric lemma first mentionned in \cite{Brunet07PLA}.

\subsection{A Geometric Lemma}

Let us consider the  Hilbert space $\mb R^3$, and let $u$ and $v$ denote two non-nul vectors such that $u\cdot v >0$. In an appropriate orthonormal basis $\mb e=\coll{e_1,e_2,e_3}$, one can write:
$$ u = \left( \begin{array}{c} 1\\0\\0 \end{array} \right) \qquad v = \left( \begin{array}{c}
\cos \theta \\ \sin \theta \\ 0 \end{array} \right) $$
with $\theta \in \, ]0,\frac\pi2[$.

In the same basis, given a real $\varphi$, let us introduce $w_\varphi = \left(\begin{array}{c}
0 \\ \cos \varphi \\ \sin \varphi \end{array} \right) $.

Let $E_\varphi$ be the plan spanned by $u$ and $w_\varphi$: $E_\varphi=\hbox{span}\coll{u,w_\varphi}$ and let $\pi_\varphi(v)$ denote the orthogonal projection of $v$ on $E_\varphi$. Finally, let $v_\varphi$ denote this projection after normalization:

$$ v_\varphi = \frac{\pi_\varphi(v)}{||\pi_\varphi(v)||}$$

Simple calculations show that:

$$ v_\varphi=\frac1{\sqrt{\cos^2\theta+\sin^2\theta\cos^2\varphi}} \left( \begin{array}{c} \cos \theta \\ \sin \theta \cos^2 \varphi \\ \sin \theta \cos \varphi \sin \varphi \end{array} \right) $$

As a consequence, one has:
$$ v_\varphi \cdot v_\psi = \frac {\cos^2 \theta+\sin^2 \theta \paren {\cos^2 \varphi \cos^2\psi + \cos \varphi \cos \psi \sin \varphi \sin \psi } } {\sqrt{\paren{\cos^2\theta+\sin^2\theta\cos^2\varphi}\paren{\cos^2\theta+\sin^2\theta\cos^2\psi}}}$$

The study of this equality leads to:
\begin{Prop}[Geometric Lemma]
With the previous notations, if $0<\theta<\frac\pi2$, one has: 
$$ \set{v_\varphi \cdot v_\psi}{\varphi, \psi \in [0;2\pi]}=\left[\frac{3\cos\theta-1}{\cos\theta+1};1\right]$$
\end{Prop}
\begin{Proof}
It is clear that the function $\pair \varphi \psi \mapsto v_\varphi\cdot v_\psi$ is a $C^\infty$ function, so that the set on the left-hand side of the equality has to be an interval. In order to find the extremal values, we need to solve the system:
$$ \left\{\begin{array}{l} \cfrac {\partial (w_\varphi \cdot w_\psi)} {\partial \varphi} = 0 \\ \cfrac{\partial (w_\varphi \cdot w_\psi)}{\partial \psi} = 0 \end{array} \right. $$
Simple calculations show that:
\begin{align*}
& \left\{\begin{array}{l} \cfrac {\partial (w_\varphi \cdot w_\psi)} {\partial \varphi} = 0 \\ \cfrac{\partial (w_\varphi \cdot w_\psi)}{\partial \psi} = 0 \end{array} \right. \\
& \Leftrightarrow \left\{\begin{array}{l} 2 \cos^3 \varphi \cos \psi \sin^2 \theta \sin (\varphi-\psi)+\cos^2 \theta \sin 2(\varphi-\psi) = 0 \\ 2 \cos \varphi \cos^3 \psi \sin^2 \theta \sin (\varphi-\psi)+\cos^2 \theta \sin 2(\varphi-\psi) = 0 \end{array} \right. \\
& \Leftrightarrow \left\{
	\begin{array}{l}
		2 \cos^3 \varphi \cos \psi \sin^2 \theta \sin (\varphi-\psi)+\cos^2 \theta \sin 2(\varphi-\psi) = 0 \\
		\cos \varphi \cos \psi \sin (\varphi-\psi) ( \cos^2 \varphi - \cos^2 \psi) = 0
	\end{array}
\right.
\end{align*}
Now, considering the second equality, one has:
\begin{align*}
& \cos \varphi \cos \psi \sin (\varphi-\psi) ( \cos^2 \varphi - \cos^2 \psi) = 0 \\
& \Leftrightarrow \paren {\cos \varphi = 0 \cor \cos \psi = 0 \cor \sin(\varphi-\psi)=0 \cor \cos^2 \varphi = \cos^2 \psi} \\
& \Leftrightarrow \paren {\varphi \equiv \frac\pi2\,[\pi] \cor \psi \equiv \frac\pi2\,[\pi] \cor \varphi\equiv \psi\,[\pi] \cor \varphi \equiv - \psi\,[\pi]}
\end{align*}
If $\varphi \equiv \psi \, [\pi]$, then $v_\varphi \cdot v_\psi = 1$.
If $\varphi \equiv \cfrac \pi 2 \, [\pi]$ or $\psi \equiv \cfrac \pi 2\, [\pi]$, the first equality yields:
\begin{align*}
& 2 \cos^3 \varphi \cos \psi \sin^2 \theta \sin (\varphi-\psi)+\cos^2 \theta \sin 2(\varphi-\psi) = 0 \\
& \Leftrightarrow \sin 2(\varphi-\psi) = 0 \\
& \Leftrightarrow \varphi \equiv \psi \, [\frac\pi2]
\end{align*}
In that case, $v_\varphi \cdot v_\psi $ equals either $1$ or $\cos \theta$. Finally, if $\varphi \equiv - \psi \, [\pi]$, then:
\begin{align*}
& 2 \cos^3 \varphi \cos \psi \sin^2 \theta \sin (\varphi-\psi)+\cos^2 \theta \sin 2(\varphi-\psi) = 0 \\
& \Leftrightarrow 2 \cos^4 \varphi \sin^2\theta \sin 2 \varphi + \cos^2 \theta \sin 4 \varphi = 0 \\
& \Leftrightarrow \sin \varphi \cos \varphi \paren { \sin^2 \theta \cos^4 \varphi + 2 \cos^2 \theta \cos^2 \varphi - \cos^2 \theta} = 0 \\
& \Leftrightarrow \paren { \sin \varphi = 0 \cor \cos \varphi = 0 \cor \sin^2 \theta \cos^4 \varphi + 2 \cos^2 \theta \cos^2 \varphi - \cos^2 \theta = 0}
\end{align*}
In the third case, the equation $\sin^2 \theta X^2 + 2 \cos^2 \theta X - \cos^2 \theta = 0$ has a single positive solution: $\displaystyle X = \frac{-2\cos^2\theta+2\cos\theta}{2\sin^2\theta} = \frac{\cos \theta}{1+\cos\theta}$. Thus, if $\varphi \equiv - \psi \, [\pi]$, the first equality becomes:
\begin{align*}
& 2 \cos^3 \varphi \cos \psi \sin^2 \theta \sin (\varphi-\psi)+\cos^2 \theta \sin 2(\varphi-\psi) = 0 \\
& \Leftrightarrow \paren { \sin \varphi = 0 \cor \cos \varphi = 0 \cor \cos^2 \varphi = {\frac{\cos \theta}{1+\cos\theta}}}
\end{align*}
And consequently, $v_\varphi \cdot v_\psi$ equals either $1$ (if $\cos \varphi=0$ or $\sin \varphi=0$) or $\cfrac {3 \cos \theta - 1}{\cos \theta + 1} $ (if $\cos^2\varphi = \cfrac{\cos \theta}{1+\cos\theta}$).

To sum things up, the values of $v_\varphi \cdot v_\psi$ obtained when $\cfrac {\partial (w_\varphi \cdot w_\psi)} {\partial \varphi} = 0$ and $\cfrac{\partial (w_\varphi \cdot w_\psi)}{\partial \psi} = 0$ are $1$, $\cos \theta$ and $\cfrac {3 \cos \theta - 1}{\cos \theta + 1}$. We conclude by remarking that $\cfrac {3 \cos \theta - 1}{\cos \theta + 1} \leq \cos \theta \leq 1$.
\end{Proof}

Let us define $f$ on $[0,1]$ by $f(x)=\frac{3x-1}{x+1}$. It is easy to verify that $f(0)=-1$, $f(\frac13)=0$, $f(1)=1$, $f$ is strictly increasing and that $f(x)<x$. Moreover, let us define a sequence $(c_n)$ by:
$$ \left\{\begin{array}{l} c_0 = 0 \\ c_{n+1} = f^{-1}(c_n) = \cfrac{1+c_n}{3-c_n} \end{array} \right. $$
From its definition, it appears that $(c_n)$ is an homographic sequence and one can express $c_n$ as a function of $n$:
$$\fall {n \in \mb N} c_n = \cfrac n {n+2}\hbox{.}$$
Finally, we define $(\theta_n)$ by:
$$\fall {n \in \mb N} \theta_n = \arccos\paren{c_n} = \arccos\Bigl(\cfrac n {n+2}\,\Bigr)\hbox{.}$$
Clearly, $\theta_0=\frac\pi2$ and $\lim\limits_{n\rightarrow\infty} \theta_n=0$.

\subsection{The Topology of $\mc P_{\mc H}$ from a Lattice Point of View}

It is a common result that given a Hilbert space, its projective space $\mc P_{\mc H}$ is a metric space, hence it is a topological space. Its metrics is given by:
$$ \fall {A,B \in \mc P_{\mc H}} d(A,B)=\arccos\frac{|u\cdot v|}{||u||\,||v||} $$
where $u$ and $v$ are vectors of $\mc H$ such that $A=\hbox{span}(u)$ and $B=\hbox{span}(v)$. % This metrics can be extended to the Hilbert lattice $\mc L_{\mc H}$ of closed subspaces of $\mc H$ by:

Thus, the metrics (and the topology) of $\mc P_{\mc H}$ can be directly derivated from the metrics of $\mc H$.

% @@@ Extend to a metrics on $\mc L_{\mc H}$?

\medskip

The previous lemma suggests another way to define the topology of $\mc P_{\mc H}$ out of the structure of the Hilbert lattice $\mc L_{\mc H}$ of all closed subspaces of $\mc H$. In order to present this, let us first introduce a few definitions.

Given a vector $v \in \mc H$ and a closed subspace $E \in \mc L_{\mc H}$, let $\Pi_E(v)$ denote the orthogonal projection of $v$ on $E$. Now, given two closed subspaces $E$ and $F$, let $E \,\&\, F$ denote the closed subspace obtained as the union of the projection on $F$ of all vectors of $E$:
$$ E\,\&\,F = \set {\Pi_F(v)}{v \in E} $$

\begin{Lemma}\label{Lemma:Induction}
Let $\mc H$ be a Hilbert space of dimension at least $3$. Let $A$ and $B$ be in $\mc P_{\mc H}$ and $n$ an integer such that $d(A,B) \geq \theta_n$. Then there exists two closed subspaces $A_1$ and $A_2$ greater than $A$, such that $d(B\,\&\,A_1,B\,\&\,A_2)=\theta_{n-1}$.
\end{Lemma}
\begin{Proof}
Let $u$ and $u$ be two normalized vectors of $\mc H$ such that:
$$ A = \hbox{span}\{u\} \qquad B = \hbox{span}\{v\} \qquad u\cdot v = \cos d(A,B) $$
Using previous notations, it is possible to define an orthonormal basis $\mb e=\coll{e_1; e_2; e_3; \ldots}$ of $\mc H$ such that:
$$ u = e_1 \qquad v = \cos d(A,B) \, e_1 + \sin d(A,B) \, e_2 $$
Now, for every $\varphi \in \mb R$, let us define as previously: 
$$w_\varphi = \cos \varphi \, e_2 + \sin \varphi \, e_3 \qquad E_\varphi = \hbox{span}\{u,w_\varphi\} \qquad v_\varphi = \cfrac{\Pi_{E_\varphi}(v)}{||\Pi_{E_\varphi}(v)||} $$
Then, $\hbox{span}\{v_\varphi\} = \hbox{span}\{\Pi_{E_\varphi}(v)\} = \hbox{span}\{v\}\,\&\,E_\varphi = B \,\&\, E_\varphi$, with $A \subseteq E_\varphi$. From the hypothesis that $d(A,B) \geq \theta_n$ (or equivalently that $\cos d(A,B) \leq \cos(\theta_n)$) and using the monotony of $f$, one~has:
$$ \cfrac{3\cos d(A,B)-1}{\cos d(A,B)+1} = f\bigl(\cos d(A,B)\bigr) \leq f\bigl(\cos \theta_n) = \cos \theta_{n-1} $$
Using our Geometric Lemma, this implies that there exists two real numbers $\alpha$ and $\beta$ such that $v_\alpha \cdot v_\beta = \cos \theta_{n-1}$.

Finally, with $A_1=E_\alpha$ and $A_2=E_\beta$, we get the expected result, as $A \subseteq A_1$, $A \subseteq A_2$ and $d(B\,\&\,A_1,B\,\&\,A_2)=d\bigl(\hbox{span}(v_\alpha),\hbox{span}(v_\beta)\bigr)=\theta_{n-1}$.
\end{Proof}

\begin{Prop} \label{Prop:HilbertInduction}
Let $\mc H$ be a Hilbert space of dimension at least $3$. For $A$ and $B$ in $\mc P_{\mc H}$ and $n$ an integer, one has:
\begin{align*}
& d(A,B) \geq \theta_n \Leftrightarrow \\ & \quad \fexist {A_0, B_0, \ldots, A_n, B_n \in \mc P_{\mc H}} \left\{\begin{array}{l} A_n = A \cand B_n = B \\ A_k, B_k \in \&^\#(A_{k+1}, B_{k+1}) \\ A_0 \subseteq B_0^\bot \end{array}\right.
\end{align*}
where $\&^\#(A,B)=\set{B\,\&\,A'}{A' \in \mc L_{\mc H} \cand A \subseteq A'}$.
\end{Prop}
\begin{Proof}
This follows from an induction on $n$. If $n=0$, one has $d(A,B)=\frac\pi2 \Leftrightarrow A \subseteq B^\bot$. For $n \geq 1$, using lemma \ref{Lemma:Induction}, if $d(A,B) \geq \theta_n$, then there exists $A_{n-1}$ and $B_{n-1}$ in $\&^\#(A_n, B_n)$ such that $d(A_{n-1}, B_{n-1}) = \theta_{n-1}$.
\end{Proof}

Proposition \ref{Prop:HilbertInduction} shows that in dimension at least $3$, for all $A \in \mc P_{\mc H}$, the open ball $\mc B(A,\theta_n)$ can be defined without any reference to the metric space of $\mc P_{\mc H}$ and thus to $\mc H$. Instead, it can be defined using solely the structure of $\mc L_{\mc H}$. Moreover, since $\lim\limits_{n \rightarrow \infty} \theta_n=0$, the collection $\coll{\mc B(A,\theta_n)}_{n \in \mb N}$ is a neighbourhood base of $x$. As the consequence, the usual metric-based topology of $\mc P_{\mc H}$ is equivalent to the topology generated by these neighbourhood bases.

\medskip

This discussion suggests strongly that a similar construction can be defined for generalizations of Hilbert lattice, namely orthomodular lattices, since in particular the binary operation $\&$ used to express orthogonal projection is actually a usual orthomodular lattice operation called the {\em Sasaki projection} and defined as $x \,\&\, y = y \wedge (x \vee y^\bot)$ \cite{Brunet04IQSA,DallaChiara2001QuantumLogic}. In complement to its definition, we give two important properties that the Sasaki projection verifies:
\begin{Prop}
Given an orthomodular lattice $\mc L$ and three elements $a,b,c \in \mc L$, one has:
\begin{enumerate}
\item $ a \leq b \Rightarrow a \sas c \leq b \sas c $
\item $ a \sas b \leq c \Rightarrow c^\bot \sas b \leq a^\bot$
\end{enumerate}
\end{Prop}

\section{A First Atom-Based Topology}

In this section, unless stated otherwise, $\mc L$ is an atomic orthomodular lattice which verifies the following property, where ${\rm At}(\mc L)$ denotes the set of atoms of $\mc L$ and $\bot$ is the least element of $\mc L$:
$$ \fall {\pair a b \in {\rm At}(\mc L) \times \mc L} a \,\&\, b \in {\rm At}(\mc L) \cup \coll \bot \eqno{\hbox{\bf Atom Projection}} $$

This property states that the Sasaki projection of a atom on any element of $\mc L$ is either $\bot$ or an atom. It is easy to verify that if $\mc H$ is a Hilbert space, then $\mc L_{\mc H}$ is atomic and verifies this property.

% It is know that this property is sufficient to ensure that the lattice $\mc L$ verifies the covering law property (references ?)

\subsection{A Topology on Atoms}

Let us first define the following sequence of subsets of ${\rm At}(\mc L)^2$:
\begin{gather*}
R^{At}_0=\set{\pair ab \in {\rm At}(\mc L)^2}{a \leq b^\bot} \\
R^{At}_{n+1} = \set{\pair ab \in {\rm At}(\mc L)^2}{\fexist {\pair{a'}{b'} \geq \pair a b} \pair {a \,\&\, b'}{b \,\&\, a'} \in R^{At}_n}
\end{gather*}
Intuitively, $\pair a b$ is in $R^{At}_n$ if and only if two orthogonal atoms can be reached in $n$ steps similar to that of proposition \ref{Prop:HilbertInduction}. And for a Hilbert space, this corresponds to $d(A,B) \geq \theta_n$, as we will soon show.

\begin{Prop} \label{Prop:EasyFacts1}
Here are some easy facts about $\coll{R^{At}_n}$, with $n \in \mb N$ and $a,b \in {\rm At}(\mc L)$:
$$ R^{At}_n \subseteq R^{At}_{n+1} \qquad  \pair a a \not \in R^{At}_n \qquad \pair a b \in R^{At}_n \Leftrightarrow \pair b a \in R^{At}_n $$
\end{Prop}
\begin{Proof}
First, for all $x$, one has $x \leq \top$ and thus $x \,\&\, \top =x$, so that if $\pair a b$ is in $R^{At}_n$, then $ \pair a b = \pair {a \,\&\, \top}{b \,\&\, \top} \in R^{At}_{n+1}$. To show the second point, for $n \geq 1$, if $\pair a a \in R^{At}_n$ then it is easy to show that $\pair a a \in R^{At}_{n-1}$ and, by induction, that $\pair a a \in R^{At}_0$ which is not possible since $a \not \leq a^\bot$. The third point follows directly from the fact that it holds for $R^{At}_0$ and from the symmetry of the definition.
\end{Proof}

The collection $\coll{R^{At}_n}$ specifies how much outspread atoms are. Furthermore, we define:
\begin{gather*}
R^{At}_\infty = \bigcup\limits_{n=0}^\infty R^{At}_n \\
\fall {a \in {\rm At}(\mc L)} \fall {n \in \mb N} B^{At}_n(a)  = \set {b \in {\rm At}(\mc L)}{\pair a b \in R^{At}_\infty \setminus R^{At}_n}
\end{gather*}
These subsets $B^{At}_n(a)$ will play a role similar to “open balls” in our topology. Here are a few facts following directly from the definitions and from proposition~\ref{Prop:EasyFacts1}:
\begin{Prop}
If $n_1 \leq n_2$ then $B^{At}_{n_2}(a) \subseteq B^{At}_{n_1}(a)$. Moreover, one has $b \in B^{At}_n(a) \Leftrightarrow a \in B^{At}_n(b)$ and $\bigcap_n B^{At}_n(a) = \emptyset$.
\end{Prop}

We now have the elements needed to define our first, atom-based topology:

\begin{Def}[Atom-Based Topology, version 1]
Let $\mc T_{At}(\mc L)$ be defined as:
$$ \mc T_{At}(\mc L) = \set{O \subseteq {\rm At}(\mc L)}{\fall {a \in O} \fexist {n \in \mb N} B^{At}_n(a) \subseteq O} $$
\end{Def}

\begin{Prop} \label{Prop:TopAt}
The set $\mc T_{At}(\mc L)$ is a topology on ${\rm At}(\mc L)$.
\end{Prop}
\begin{Proof}
Clearly, both ${\rm At}(\mc L)$ and $\emptyset$ are in $\mc T_{At}(\mc L)$, and $\mc T_{At}(\mc L)$ is stable by arbitrary union.

Now, let $O_1$ and $O_2$ be two elements of $\mc T_{At}(\mc L)$, and let $a$ be an element of $O_1 \cap O_2$. By definition of $\mc T_{At}(\mc L)$, there exists two integers $n_1$ and $n_2$ such that $B^{At}_{n_1}(a) \subseteq O_1$ and $B^{At}_{n_2}(a) \subseteq O_2$. With $n=\max(n_1, n_2)$,  $B^{At}_n(x) = B^{At}_{n_1}(x) \cap B^{At}_{n_2}(x)$ so that $B^{At}_n(x) \subseteq O_1 \cap O_2$.

This shows that $O_1 \cap O_2$ is in $\mc T_{At}(\mc L)$.
\end{Proof}

Let us now turn to the case where $\mc L$ is a Hilbert lattice $\mc L_{\mc H}$ with $\dim \mc H \geq 3$. We first give two technical results before showing that $\mc T_{At}(\mc L_{\mc H})$ and the metrics-induced topology on $\mc P_{\mc H}$ are equivalent.
\begin{Prop} \label{Prop:CrossOML}
For any orthomodular lattice $\mc L$ and $a, b, c \in \mc L$, one has:
$$ b \sas (a \vee c) \leq c \Leftrightarrow a \sas (b \vee c) \leq c $$
\end{Prop}
\begin{Proof}
One has $ b \sas (a \vee c) \leq c \Leftrightarrow c^\bot \sas (a \vee c) \leq b^\bot $. Now, since $c^\bot$ and $c \vee a$ are compatible, $c^\bot \sas (a \vee c) = c^\bot \wedge (a \vee c) = a \sas c^\bot$. This leads to:
$$ b \sas (a \vee c) \leq c \Leftrightarrow a \sas c^\bot \leq b^\bot \Leftrightarrow b \sas c^\bot \leq a^\bot \Leftrightarrow a \sas (b \vee c) \leq c $$
\end{Proof}

\begin{Prop} \label{Prop:CrossOML2}
Let $\mc L$ be an atomic OML verifying the {\em atom-projection} property, and let $a$, $b$ and $c$ be three atoms such that $b \sas (a \vee c) = c$ and $a \not \leq b^\bot$. Then one has $a \sas (b \vee c) = c$.
\end{Prop}
\begin{Proof}
As a direct consequence of proposition \ref{Prop:CrossOML}, one has $a \sas (b \vee c) \leq c$. But having $a \sas (b \vee c) = \bot$ would mean that $a \leq (b \vee c)^\bot$ so that in particular $a \leq b^\bot$ which is not possible. Thus, the only possibility is $a \sas (b \vee c) = c$.
\end{Proof}

\begin{Prop} \label{Prop:TopoHilbert1}
Given a Hilbert space $\mc H$ with $\dim \mc H \geq 3$, the topology $\mc T_{At}(\mc L_{\mc H})$ is equivalent to the topology induced on $\mc P_{\mc H}$ by the metric of $\mc H$.
\end{Prop}
\begin{Proof} In proposition \ref{Prop:HilbertInduction}, we used the notation $\&^\#(a,b) = \set {b \sas a'}{a' \geq a}$. If $a \not \leq b^\bot$, for every $c$ in $\&^\#(a,b)$, one has $c = b \sas (a \vee c)$ so that using proposition \ref{Prop:CrossOML2}, $c \in \&^\#(b,a)$ and more generally, $\&^\#(a,b)=\&^\#(b,a)$. With this remark and proposition \ref{Prop:HilbertInduction}, it follows that for all $n \in \mb N$:
$$ R^{At}_n = \set {\pair a b}{d(a,b) \geq \theta_n} $$
Thus, for $a \in \mc P_{\mc H}$ and $n \in \mb N$, one has $B^{At}_n(a) = \set{b \in \mc P_{\mc H}}{0 < d(a,b) < \theta_n}$. And since $\lim \theta_n = 0$, it follows that:
$$ O \in \mc T_{At}(\mc L_{\mc H}) \ \Leftrightarrow \ \paren{\fall {x \in O} \fexist {\delta>0} \fall {y \in \mc P_{\mc H}} d(x,y)<\delta \Rightarrow y \in O} $$
Stated another way, $O$ is in $\mc T_{At}(\mc L_{\mc H})$ if and only if it is open with regards to the topology induced on $\mc P_{\mc H}$ by the metric of $\mc H$.
\end{Proof}

\subsection{Extending the Topology to the Entire Lattice}

In order to generalize the definition of our topology to the whole lattice, we introduce the following collection of “balls”:
$$ \fall {a \in \mc L} B^L_n(a)=\set{b \in \mc L}{\begin{array}{l} \fall {x \in {\rm At}(a)} \fexist {y \in {\rm At}(b)} y \in B^{At}_n(x) \\ \hbox{and\ } \fall {y \in {\rm At}(b)} \fexist {x \in {\rm At}(a)} x \in B^{At}_n(y) \end{array}} $$
Here, given an element $x$ of $\mc L$, ${\rm At}(x)$ denotes the set of atoms below $x$. The sets $\coll{B^L_n(a)}$ enable us to define a topology on $\mc L$ the following way.
\begin{Def}[Atom-Based Topology, version 2]
Let $\mc T_{L}(\mc L)$ be defined as:
$$ \mc T_{L}(\mc L) = \set{O  \subseteq \mc L}{\fall {a \in O} \fexist {n \in \mb N} B^L_n(a) \subseteq O} $$
\end{Def}

\begin{Prop}\label{Prop:TopLat}
The set $\mc T_{L}(\mc L)$ is a topology on $\mc L$.
\end{Prop}
\begin{Proof}
The proof is similar to that of proposition \ref{Prop:TopAt}.
\end{Proof}

The next proposition shows that the first topology, defined on atoms, can be obtained from the topology on the whole lattice:

\begin{Prop}
One has $\mc T_{At}(\mc L) = \set{O \cap {\rm At}(\mc L)}{O \in \mc T_{L}(\mc L)} $.
\end{Prop}
\begin{Proof}
This follows from the definition of $B^L_n(a)$ and the fact that if $a$ is an atom, then ${\rm At}(a)=\coll a$.
\end{Proof}

We now turn once again to the case where the considered orthomodular lattice is actually a Hilbert lattice $\mc L_{\mc H}$. The metrics induced on $\mc P_H$ by the metrics of $\mc H$ can be generalized on $\mc L_{\mc H}$ by defining:
$$ d_L(a,b) = \max \paren{ \max_{x \in {\rm At}(a)} d_\pi(x, b), \max_{y \in {\rm At}(b)} d_\pi(y, a)} $$
where for $\pair x b \in \mc P_{\mc H} \times \mc L_{\mc H}$, $d_\pi(x,b) = \min\limits_{y \in {\rm At}(b)} d(x,y)$.
\begin{Prop} \label{Prop:TopoHilbert2}
Given a Hilbert lattice $\mc L_{\mc H}$ where $\dim \mc H \geq 3$, the topology $\mc T_L(\mc L_{\mc H})$ is equivalent to the topology induced on $\mc L_{\mc H}$ by the metrics of $\mc H$.
\end{Prop}
\begin{Proof}
This is a direct consequence of the definition of the metrics on $\mc L_{\mc H}$ and of proposition \ref{Prop:TopoHilbert1} where we have shown that $B^{At}_n(a) = \set {b \in \mc P_{\mc H}}{0<d(a,b)<\theta_n}$. In order to show a similar result for $B^L_n(a)$, we have to consider two possibilities:
\begin{enumerate}
\item If $a$ is an atom, we have $B^L_n(a) = \set {b \in \mc L}{\fall {y \in {\rm At}(b)} 0< d(a, y) <\theta_n}$. In that case, if $b$ is in $B^L_n(a)$, $b$ has to be an atom too (otherwise, $b \wedge a^\bot \neq \bot$ and there exists an atom $y$ of $b$ such that $d(a,y)=\frac\pi2$). Thus, we have:
$$ B^L_n(a) = \set {b \in {\rm At}(\mc L)}{0<d(a,b)<\theta_n} = \set {b \in \mc L}{0 < d_L(a,b) < \theta_n} $$
\item If $a$ is not an atom, suppose that $b$ is in $B^L_n(a)$. From the discussion for the previous case, $b$ cannot be an atom. And following the definition of $B^L_n(a)$, one has:
$$ \begin{array}{l} \fall {x \in {\rm At}(a)} \fexist {y \in {\rm At}(b)} 0< d(x,y) <\theta_n \\ \hbox{and\ } \fall {y \in {\rm At}(b)} \fexist {x \in {\rm At}(a)} 0< d(x,y) <\theta_n \end{array} $$
For $x \in {\rm At}(a)$, two subcases have to be considered. If $x \not \in {\rm At}(b)$, then we have:
$$ \paren{\fexist{y \in {\rm At}(b)} 0 < d(x,y) < \theta_n} \ \Leftrightarrow \ 0 < d_\pi(x,b) < \theta_n $$
And if $x \in {\rm At}(b)$, then the condition $\fexist {y \in {\rm At}(b)} 0 < d(x,y) <\theta_n$ is always verified since $\dim b \geq 2$. This implies that if $a$ is not an atom, then:
\begin{align*}
B^L_n(a) & = \set {b \in {\mc L}}{\begin{array}{l} \fall {x \in {\rm At}(a)\setminus {\rm At}(b)} 0 < d_\pi(x,b) <\theta_n \\ \hbox{and\ } \fall {y \in {\rm At}(b) \setminus {\rm At}(a)} 0 < d_\pi(x,a) <\theta_n \end{array}} \\
& = \set {b \in {\mc L}}{\begin{array}{l} \fall {x \in {\rm At}(a)} 0 \leq d_\pi(x,b) <\theta_n \\ \hbox{and\ } \fall {y \in {\rm At}(b)} 0 \leq d_\pi(x,a) <\theta_n \end{array}} \\
& = \set {b  \in {\mc L}}{0 \leq d_L(a,b) < \theta_n}
\end{align*}
\end{enumerate}
Thus, for a subset $S \subseteq \mc L$ and an element $a \in S$, then the following two statements $\fexist {\theta>0}{\fall b (d_L(a,b)<\theta \Rightarrow b \in S)}$ and $\fexist n B_n(a) \subseteq S$ are equivalent. As a conclusion, the two topologies are equivalent.
\end{Proof}

\section{The General Construction}

In order to extend our topology to the general case, we first need to adapt the needed definitions in order to avoid any references to atoms. This is done using the tools described in the next section.

\subsection{Operations on Lower Sets}

We recall that given poset $P$, a subset $I \subseteq P$ is called a {\em lower set} if and only if it verifies:
$$ \fall {x \in I} \fall {y \leq x} y \in I $$
Let $\wp^\downarrow(P)$ denote the lower sets of a poset $P$. In the following, we introduce a complementation and a closure operation on lower sets which will enable us to avoid direct reference to atoms.

\begin{Def}[Complement of a Lower Set]
For $I \in \wp^\downarrow(P)$, we define its complement $\neg I$ as:
$$ \neg I = \set {x \in P}{\fall {y \leq x} (y \in I \Rightarrow y=\bot)} $$
\end{Def}
% The binary relation $\leq_\star$ is a partial order on $P$ which “disconnects” $\bot$ from the rest.
We first present some easy fact about this operation:
\begin{Prop}
For $I \in \wp^\downarrow(P)$, one has:
$$ \neg I \in \wp^\downarrow(P) \quad \quad I \subseteq \neg \neg I \quad \quad \neg \neg \neg I = \neg I \quad \quad I \cap \neg I = \coll \bot $$
\end{Prop}
\begin{Cor}
The operation $I \mapsto \neg \neg I$ is a closure operator on $\wp^\downarrow(P)$.
\end{Cor}
This closure operator is not new and appears for instance in the domains of Kripke's possible  worlds semantics of modal and intuitionnistic logic. In intuitionnistic logic, if a proposition $p$ is interpreted by an ideal $I$, then its double-negation $\neg\neg p$ is interpreted by $\neg \neg I$ \cite{VanDalen86Handbook}. Similarly, in S4 modal logic, $\neg\neg I$ is the interpretation of $\Box\Diamond p$ \cite{Chellas80Book,HughesCresswell96Book}.
In both case, the intuitive idea is the same: an element $x$ is in the closure $\neg\neg I$ if and only if every path starting from $x$ downwards ends in $I$. This appears clearly in the following expression of $\neg \neg I$: 
$$ \neg \neg I = \set x {\fall {y \leq x} ((\fexist{z \leq y} z \in I \setminus \coll \bot) \vee y = \bot)}\hbox{,} $$

\begin{Prop}
Given an atomic poset $P$, a ideal $I \in \wp^\downarrow(P)$ and an element $x \in P$, one has:
$$ x \in \neg \neg I \ \Leftrightarrow \fall {y \in {\rm At}(x)} y \in I $$
\end{Prop}
\begin{Proof}
First, suppose that $x$ is in $\neg \neg I$ and let $y$ be in ${\rm At}(x)$. Since  $y \leq x$ and $y \neq \bot$, there exists a $z$ such that $z \leq y$ and $z \in I \setminus \coll \bot$. But since $y$ is an atom, one has $z=y$ and thus $y \in I$. We have shown that $\fall {y \in {\rm At}(x)} y \in I$.

Conversely, let $x$ be in $P$ such that $x \neq \bot$ and $\fall {z \in {\rm At}(x)} z \in I$ and let $y$ be such that $\bot < y \leq x$. For all $z$ be in ${\rm At}(y)$, we have $z \in {\rm At}(x)$ and thus $z \in I$. As a consequence, we have shown that $\fall {y \leq x} y \neq \bot \Rightarrow \fexist {z \leq y} z \in I \setminus \coll \bot$.
\end{Proof}
\begin{Cor} \label{Cor:Closure}
Let $I$ be an ideal of an atomic poset $P$. If ${\rm At}(P)$ denotes the atoms of $P$, then one has:
$$ \neg \neg I \cap {\rm At}(P) = I \cap {\rm At}(P) $$
\end{Cor}

This proposition and its corollary show that the complement operation and the derived closure operation provide an interesting basis for generalizing the definition of our topology.

\begin{Prop} \label{Prop:AtomComplement}
Given a poset $P$, a lower set $I \in \wp^\downarrow(P)$ and an atom $x$ of $P$, one has:
$$ x \in \neg I \ \Leftrightarrow x \not \in I $$
\end{Prop}

\subsection{Definition of the Topology in the General Case}

Before turning to the definition of our topology, we introduce the smashed-product of a poset which will avoid excessive trouble when dealing with $\bot$.

\begin{Def}[Smashed-product]
Given a poset $P$ with a least element $\bot$, we define the $\bot$-smashed-product $P^{\#2}$ as the poset with elements $\set{\pair a b \in P^2}{a \neq \bot \cand b \neq \bot} \cup \coll {\pair \bot \bot}$ equip\-ped with the point-wise partial order induced by $P$.
\end{Def}

Obviously, if $P$ is atomic, then so is $P^{\#2}$, in which case ${\rm At}(P^{\#2}) = {\rm At}(P) \times {\rm At}(P)$. In the following, we will consider the $\bot$-smashed-product $\mc L^{\#2}$ of an orthomodular lattice $\mc L$. Of course, $\mc L^{\#2}$ is a poset (and even a lattice) but not an orthomodular lattice.

\medskip

We now have the tools to adapt our topology-related definitions to the general case. 

\begin{Def}[Topology]
Given an orthomodular lattice $\mc L$, we first define a collection $\coll{R_n}$ telling how “far apart” elements of $\mc L$ are:
\begin{gather*}
R_0 = \neg \neg {\set {\pair a b \in \mc L^{\#2}}{a \leq b^\bot}} \\
R_{n+1} = \neg \neg {\set {\pair a b \in \mc L^{\#2}}{\fexist{\pair {a'}{b'}\geq \pair a b} \pair{a\,\&\,b'}{b\,\&\,a'} \in R_n}} 
\end{gather*}
Then, we introduce subsets $B_n(a)$ specifying which elements are close enough from $a$:
\begin{gather*}
R_\infty = \bigcup_{k=0}^\infty R_k \\
B_n(a) = \set {b \in \mc L} {\begin{array}{l} \fall {a' \leq_{\not\bot} a} \fexist {a'' \leq_{\not\bot} a'} \fexist {b' \leq_{\not\bot} b} \pair {a''}{b'} \in R_\infty \cap \neg R_n \\ \hbox{and\ }\fall {b' \leq_{\not\bot} b} \fexist {b'' \leq_{\not\bot} b'} \fexist {a' \leq_{\not\bot} a} \pair {a'}{b''} \in R_\infty \cap \neg R_n \end{array} }
\end{gather*}
In this definition, $x \leq_{\not\bot} y$ means $\bot < x \leq y$.
Finally, we define the topology $\mc T(\mc L)$:
$$ \mc T(\mc L) = \set{O  \subseteq \mc L}{\fall {a \in O} \fexist {n \in \mb N} B_n(a) \subseteq O} $$
\end{Def}

Some remarks have to be done concerning the definition of $\coll{R_n}$. First, it is clear that $\set {\pair a b \in \mc L^{\#2}}{a \leq b^\bot}$ is a lower set of $\mc L^{\#2}$ so that the definition of $R_0$ makes sense. Similarly, for $I \in \wp^\downarrow(\mc L^{\#2})$, the set $J$ defined as:
$$J = \set {\pair a b \in \mc L^2}{\fexist{\pair {a'}{b'}\geq \pair a b} \pair{a\,\&\,b'}{b\,\&\,a'} \in I} $$ is also a lower set. To show this, let $\pair a b$ be in $J$ and $\pair \alpha \beta$ be such that $\pair \alpha \beta \leq \pair a b$. Since $\pair a b \in J$, there exists $\pair {a'}{b'} \geq \pair a b$ such that $\pair{a \sas b'}{b \sas a'}$ is in $I$. But then, $\pair {a'}{b'} \geq \pair \alpha \beta$ and $\pair {\alpha \sas b'}{\beta \sas a'} \leq \pair {a \sas b'}{b \sas a'}$. As a consequence, $\pair {\alpha \sas b'}{\beta \sas a'}$ is in $I$ and thus $\pair \alpha \beta$ is in $J$. This shows that $J$ is an ideal, and that the definition of $\coll{R_n}$ by induction makes sense.

\medskip

Of course, we have:

\begin{Prop}
Given an orthomodular lattice $\mc L$, $\mc T(\mc L)$ is a topology on $\mc L$.
\end{Prop}
\begin{Proof}
Once again, the proof is the same as that of proposition \ref{Prop:TopAt}.
\end{Proof}

\subsection{Equivalence of Topologies}

\begin{Prop} \label{Prop:AtomEquiv}
For all $n \in \mb N$, one has $R_n \cap {\rm At}(\mc L^{\#2}) = R_n^{At}$.
\end{Prop}
\begin{Proof}
We show this by induction. It is clearly true for $n=0$. Now, suppose that it is true for $n$ and let $a$ and $b$ be two atoms. As a consequence of corollary \ref{Cor:Closure}, one has:
$$ \pair a b \in R_{n+1} \ \Leftrightarrow\ \fexist {\pair {a'}{b'} \geq \pair a b} \pair {a \,\&\, b'}{b \,\&\, a'} \in R_n $$
As a consequence, two possibilities have to be considered:
\begin{enumerate}
\item If $a \leq b^\bot$, then $\pair a b$ is in both $R_0 \cap {\rm At}(\mc L^{\#2})$ and $R^{At}_0$ and thus, it is in both $R_{n+1} \cap {\rm At}(\mc L^{\#2})$ and $R^{At}_{n+1}$. 
\item Otherwise, for all $b' \geq b$ and $a' \geq a$, neither $a\,\&\,b'$ nor $b\,\&\,a'$ are equal to $\bot$, so that they are both atoms due to the atom projection property, and by induction:
$$\pair {a \,\&\, b'}{b \,\&\, a'} \in R_n \cap {\rm At}(\mc L^{\#2}) \ \Leftrightarrow \pair {a \,\&\, b'}{b \,\&\, a'} \in R^{At}_n$$
One then has:
\begin{align*}
\pair a b \in & \, R_{n+1} \cap {\rm At}(\mc L^{\#2}) \\ & \Leftrightarrow\ \fexist {\pair {a'}{b'} \geq \pair a b} \pair {a \,\&\, b'}{b \,\&\, a'} \in R_n \cap {\rm At}(\mc L^{\#2}) \\ & \Leftrightarrow\ \fexist {\pair {a'}{b'} \geq \pair a b} \pair {a \,\&\, b'}{b \,\&\, a'} \in R^{At}_n \\ & \Leftrightarrow\ \pair a b \in R^{At}_{n+1}
\end{align*}
\end{enumerate}
As a result, the equality $R_n \cap {\rm At}(\mc L^{\#2}) = R_n^{At}$ holds for all $n \in \mb N$.
\end{Proof}

\begin{Prop} \label{Prop:EquivTopo}
If $\mc L$ is atomic and verifies the atom projection property, then:
$$ \fall {a,b \in \mc L} \paren{b \in B_n(a) \,\Leftrightarrow\, b \in B^L_n(a)} $$
\end{Prop}
\begin{Proof}
We only have to show the following equivalence:
\begin{align*}
& \paren{\fall {a' \leq_{\not\bot} a} \fexist {a'' \leq_{\not\bot} a'} \fexist {b' \leq_{\not\bot} b} \pair {a''}{b'} \in R_\infty \cap \neg R_n} \\ & \Leftrightarrow \ \paren{\fall {x \in {\rm At}(a)} \fexist {y \in {\rm At}(b)} \pair {x} {y} \in R^{At}_\infty\setminus R^{At}_n}
\end{align*}
The $\Leftarrow$-direction is direct. Conversely, suppose that one has:
$$\fall {a' \leq_{\not\bot} a} \fexist {a'' \leq_{\not\bot} a'} \fexist {b' \leq_{\not\bot} b} \pair {a''}{b'} \in R_\infty \cap \neg R_n$$
and let $x$ be an atom of $a$. Our assumption implies that there exists an element $b' \leq_{\not\bot} b$ such that $\pair x {b'} \in R_\infty \cap \neg R_n$. Moreover, $R_\infty \cap \neg R_n$ is a lower set so that there exists an atom $y$ of $b$ such that $\pair x y \in R_\infty \cap \neg R_n$. We conclude by remarking that from proposition \ref{Prop:AtomEquiv}, one has $R_\infty \cap {\rm At}(\mc L^{\#2}) = R^{At}_\infty$ and that following proposition \ref{Prop:AtomComplement}, $\pair x y \in \neg R_n \Leftrightarrow \pair x y \not \in R_n \Leftrightarrow \pair x y \not \in R^{At}_n$.
\end{Proof}

\begin{Cor} \label{Cor:EquivTopo}
Given an atomic orthomodular lattice $\mc L$ verifying the atom projection property, the topologies $\mc T(\mc L)$ and $\mc T_L(\mc L)$ are equivalent.
\end{Cor}
\begin{Proof}
This a direct consequence of proposition \ref{Prop:EquivTopo}.
\end{Proof}

\begin{Theo} \label{Theo:TopoHilbert3}
Given an Hilbert lattice $\mc L_{\mc H}$ where $\dim \mc H \geq 3$, the topology $\mc T(\mc L_{\mc H})$ is equivalent to the topology induced on $\mc L_{\mc H}$ by the metrics of $\mc H$.
\end{Theo}
\begin{Proof}
This follows from corollary \ref{Cor:EquivTopo} and proposition \ref{Prop:TopoHilbert2}.
\end{Proof}

\section{Quantum and Classical Structures}

As we have seen in theorem \ref{Theo:TopoHilbert3}, if $\mc L$ is a Hilbert lattice $\mc L_{\mc H}$ where $\dim \mc H \geq 3$, then the topology $\mc T(\mc L_{\mc H})$ is equivalent to the one induced by the metrics of $\mc H$. In particular, with this topology, there are no isolated points except the extremal ones, i.e. $\top$ and $\bot$.

We now explore the topology $\mc T(\mc B)$ of a boolean algebra $\mc B$. First, following our definitions, $R_0 = \neg \neg \set {\pair a b}{a \leq b^\bot}$. Since $\mc B$ is a boolean algebra, one has $a \leq b^\bot \Leftrightarrow a \wedge b = \bot$ and thus:
\begin{align*}
\set {\pair a b}{a \leq b^\bot} & = \set {\pair a b}{a \wedge b = \bot} \\
\neg \set {\pair a b}{a \leq b^\bot} & = \set {\pair a b}{\fall {\pair{a'}{b'} \leq \pair a b} (a' \wedge b' = \bot \Rightarrow a' = b' = \bot)} \\
& = \set {\pair a a}{a \in {\rm At}(\mc B)} \cup \coll {\pair \bot \bot} \\
\neg \neg \set {\pair a b}{a \leq b^\bot} & = \neg \paren { \set {\pair a a}{a \in {\rm At}(\mc B)} \cup \coll {\pair \bot \bot} } \\
& = \set {\pair a b}{\fall {\pair{a'}{b'} \leq \pair a b} \pair {a'}{b'} \not \in \set {\pair a a}{a \in {\rm At}(\mc B)}} \\
& = \set {\pair a b}{{\rm At}(a) \cap {\rm At}(b) = \emptyset}
\end{align*}
We thus have $R_0 = \set {\pair a b}{{\rm At}(a) \cap {\rm At}(b) = \emptyset}$. Now, in order to compute $R_1$, we have:
\begin{align*}
& \fexist {\pair {a'}{b'} \geq \pair a b} \pair {a \sas b'}{b \sas a'} \in R_0 \\ & \quad \Leftrightarrow  \ \fexist {\pair {a'}{b'} \geq \pair a b} \pair {a \wedge b'}{b \wedge a'} \in R_0 \\ & \quad \Leftrightarrow  \ \fexist {\pair {a'}{b'} \geq \pair a b} {\rm At}(a \wedge b') \cap {\rm At}(b \wedge a') = \emptyset \\
& \quad \Leftrightarrow \ \fexist {\pair {a'}{b'} \geq \pair a b} {\rm At}(a) \cap {\rm At}(b') \cap {\rm At}(b) \cap {\rm At}(a') = \emptyset \\
& \quad \Leftrightarrow \ {\rm At}(a) \cap {\rm At}(b) = \emptyset
\end{align*}
Thus, we have $R_1 = \neg \neg R_0 = R_0$ and more generally, $R_\infty = R_n = R_0$. As a consequence, since for all $n$, $R_0 \cap \neg R_n = \coll{\pair \bot \bot}$, this implies that for $a \in \mc B$, one has $B_n(a) = \emptyset$. We have shown:

\begin{Theo} \label{Theo:Bool}
Given a boolean algebra $\mc B$, the topology $\mc T(\mc B)$ is the discrete topology on $\mc B$, that is:
$$ \mc T(\mc B) = \wp(\mc B)$$
\end{Theo}

This result shows that the topological construction we have defined on orthomodular lattices is an interesting tool for discriminating some quantum structures from classical boolean structures, since for the topology associated to a boolean algebra is discrete while the topology associated to a Hilbert lattice has only $\top$ and $\bot$ as isolated points.

\section{Conclusion and Perspective}

In this article, we have shown that it is possible to equip any orthomodular lattice with a topology. Originally, orthomodular lattices were introduced as a generalization of Hilbert lattices which are intrisically topological objects. With the construction that we provide, this suggests that orthomodular lattices can themselves be seen as intrisically topological objects.

The generality of this result is rather striking, since for the case of Hilbert lattices, the topology is induced by the underlying Hilbert space, which is strongly related to mathematical analysis. On the contrary, orthomodular lattices are purely algebraic objects and the mere structure of these objects, in particular their orthomodularity and the properties of the Sasaki projection that derive from it, are sufficient to capture their topological structure. This result can also be related to others,  such that Piron's theorem \cite{piron64} and Solèr's theorem \cite{Soler95,DallaChiara2001QuantumLogic}, which explore the way some properties of Hilbert lattices can be expressed in a lattice-theoretical way. 

However, we have just given the definition of this topology and shown a few results in order to motivate a more general study of orthomodular lattices in the light of this topology. Here are some questions for further work:
\begin{itemize}
\item Which are the {\em clopen} subsets of $\mc L$ with regards to $\mc T(\mc L)$? And in the case of a Hilbert lattice $\mc L_{\mc H}$, should the elements of $B_n(a)$ have the same dimension as $a$?
\item Under which conditions on $\mc L$ is the topology $\mc L(\mc T)$ different from the discrete one? More generally, which points are isolated in this topology. Theorems \ref{Theo:TopoHilbert3} and \ref{Theo:Bool} suggest more particularly that there exists some connection between isolated points and elements of the center of $\mc L$.
\item Which functions are continuous (in particular, we think of orthocomplementation, conjunction, the Sasaki projection, OML morphisms)?
\item In the definition of $\mc L(\mc T)$, $R_\infty$ is a lower set. Is it closed, and if not, what is its closure? The same question also applies to $R_\infty \cup \set{\pair a a}{a \in \mc L}$.
\item One can define a notion of Cauchy sequence by stating for a sequence $\coll{u_n}$ that $\fall m \fexist {\pair a n \in \mc L \times \mb N} \fall {k \geq n} u_k \in B_m(a)$. This permits to define a sort of topological completion of an OML. Then, how does this type of completion relate to other ways of making an orthomodular lattice complete?
\end{itemize}

%\bibliographystyle{abbrv}
%\bibliography{/Users/olivier/Library/texmf/biblio}

\end{document}